\documentclass[%
 aip,
 amsmath,amssymb,
 reprint,%
]{revtex4-1}

\usepackage{graphicx}
\usepackage{dcolumn}
\usepackage{bm}

\usepackage[utf8]{inputenc}
\usepackage[T1]{fontenc}
\usepackage{mathptmx}
\usepackage{etoolbox}
\usepackage[colorlinks=true, allcolors=blue]{hyperref}
\usepackage{ulem}

\makeatletter
\def\@email#1#2{%
 \endgroup
 \patchcmd{\titleblock@produce}
  {\frontmatter@RRAPformat}
  {\frontmatter@RRAPformat{\produce@RRAP{*#1\href{mailto:#2}{#2}}}\frontmatter@RRAPformat}
  {}{}
}%
\makeatother
\begin{document}

\preprint{AIP/123-QED}

\title{Continuously tunable uniaxial strain control of van der Waals heterostructure devices}
\author{Zhaoyu Liu}
\altaffiliation{Equal contributions.}
\altaffiliation[Correspond to:]{zhyliu@uw.edu}
\affiliation{Department of Physics, University of Washington, Seattle, WA, 98195, USA}

\author{Xuetao Ma}
\altaffiliation{Equal contributions.}
\affiliation{Department of Materials Science and Engineering, University of Washington, Seattle, WA 98195, USA}

\author{John Cenker}
 \affiliation{Department of Physics, University of Washington, Seattle, WA, 98195, USA}

\author{Jiaqi Cai} 
 \affiliation{Department of Physics, University of Washington, Seattle, WA, 98195, USA}
 
\author{Zaiyao Fei}
 \affiliation{Department of Physics, University of Washington, Seattle, WA, 98195, USA}

\author{Paul Malinowski}
\affiliation{Department of Physics, University of Washington, Seattle, WA, 98195, USA}

\author{Joshua Mutch}
\affiliation{Department of Physics, University of Washington, Seattle, WA, 98195, USA}

\author{Yuzhou Zhao}
 \affiliation{Department of Physics, University of Washington, Seattle, WA, 98195, USA}
 \affiliation{Department of Materials Science and Engineering, University of Washington, Seattle, WA 98195, USA}

\author{Kyle Hwangbo}
 \affiliation{Department of Physics, University of Washington, Seattle, WA, 98195, USA}

\author{Zhong Lin}
 \affiliation{Department of Physics, University of Washington, Seattle, WA, 98195, USA}
 
\author{Arnab Manna}
\affiliation{Department of Physics, University of Washington, Seattle, WA, 98195, USA}

\author{Jihui Yang}
 \affiliation{Department of Materials Science and Engineering, University of Washington, Seattle, WA 98195, USA}

\author{David Cobden}
 \affiliation{Department of Physics, University of Washington, Seattle, WA, 98195, USA}

\author{Xiaodong Xu}
 \affiliation{Department of Physics, University of Washington, Seattle, WA, 98195, USA}
 \affiliation{Department of Materials Science and Engineering, University of Washington, Seattle, WA 98195, USA}

\author{Matthew Yankowitz}
 \altaffiliation[Correspond to: ]{myank@uw.edu}
 \affiliation{Department of Physics, University of Washington, Seattle, WA, 98195, USA}
 \affiliation{Department of Materials Science and Engineering, University of Washington, Seattle, WA 98195, USA}

\author{Jiun-Haw Chu}
\altaffiliation[Correspond to: ]{jhchu@uw.edu}
\affiliation{Department of Physics, University of Washington, Seattle, WA, 98195, USA}

\date{\today}

\begin{abstract}
Uniaxial strain has been widely used as a powerful tool for investigating and controlling the properties of quantum materials. However, existing strain techniques have so far mostly been limited to use with bulk crystals. Although recent progress has been made in extending the application of strain to two-dimensional van der Waals (vdW) heterostructures, these techniques have been limited to optical characterization and extremely simple electrical device geometries. Here, we report a piezoelectric-based \textit{in situ} uniaxial strain technique enabling simultaneous electrical transport and optical spectroscopy characterization of dual-gated vdW heterostructure devices. Critically, our technique remains compatible with vdW heterostructure devices of arbitrary complexity fabricated on conventional silicon/silicon dioxide wafer substrates. We demonstrate a large and continuously tunable strain of up to $-0.15\%$ at millikelvin temperatures, with larger strain values also likely achievable. We quantify the strain transmission from the silicon wafer to the vdW heterostructure, and further demonstrate the ability of strain to modify the electronic properties of twisted bilayer graphene. Our technique provides a highly versatile new method for exploring the effect of uniaxial strain on both the electrical and optical properties of vdW heterostructures, and can be easily extended to include additional characterization techniques.
\end{abstract}

\maketitle

\section{\label{sec:level1}Introduction}

Strain is a powerful tool for directly manipulating the crystal lattice of materials, and consequently for tuning their electronic properties. For example, uniaxial strain can break the in-plane rotational symmetry of a lattice, potentially generating new electronic phases. Over the past decade, a powerful technique has been developed for applying continuously tunable uniaxial strain to three-dimensional bulk quantum materials using a device based on three piezoelectric stacks.\cite{hicks2014piezoelectric} This technique has been widely used to study and tune superconductivity,\cite{hicks2014strong,hicks2017peak,malinowski2020suppression,Qian2021CVS} topological phases,\cite{mutch2019evidence,jo2019WTe2} and nematicity\cite{bartlett2021relationship,sanchez2021transport,Ghini2021FeSe,Wiecki2021} in a variety of bulk quantum materials and to investigate their thermodynamic properties.\cite{li2020heat,ikeda2019ac} Over the past few years, van der Waals (vdW) heterostructures have emerged as exciting new platforms for exploring related strongly correlated and topological states in two dimensions. Flat electronic bands arise when two or more vdW sheets are properly stacked and can host a wealth of emergent states including superconductivity, nematicity, ferromagnetism, generalized Wigner crystals, and both integer and fractional Chern insulators.\cite{cao2018unconventional,yankowitz2019tuning, chen2021electrically, cao2021nematicity, fei2018ferroelectric, serlin2020intrinsic, balents2020superconductivity,andrei2020graphene,park2023observation,cao2020strange,polshyn2019large,jaoui2022quantum} However, technical constraints have so far severely limited the extension of existing strain-tuning techniques to the study of these states in vdW heterostructures. 

Although there have been a number of approaches developed for applying strain to vdW materials, so far none are compatible with standard electrical transport measurements of dual-gated vdW devices assembled on a silicon/silicon dioxide (Si/SiO$_2$) wafer substrate.\cite{bunch2007electromechanical,conley2013bandgap,lloyd2016band, deng2018strain,hou2019strain,dai2019strain,kim2023strain} Prior methods have mostly focused either on suspended vdW samples, which are mechanically fragile, or devices with limited numbers of gate and contact electrodes. In this paper, we introduce a novel method that integrates traditional vdW heterostructure device fabrication on silicon substrates with a three-piezo-stack-based strain cell.\cite{dean2010boron, kim2016van,yankowitz2019van, hicks2014piezoelectric, park2020rigid, bartlett2021relationship} We quantify the induced strain in the active layer of the vdW heterostructure by Raman spectroscopy and corroborate this value with estimates from metallic strain gauges either glued onto the piezo stack or evaporated onto the silicon wafer near the vdW device. We further characterize the strain transmission through multilayer vdW heterostructures by characterizing the strain induced hexagonal boron nitride (hBN) for different flake thicknesses and for twisted hBN multilayers. Our measurements indicate that strain is effectively transmitted from the silicon wafer into many layers of vdW flakes residing on top. Finally, we also showcase the capabilities of our strain technique by measuring the electrical transport properties of a twisted bilayer graphene (tBLG) device near the magic angle and show that strain can induce large changes in the device resistivity. Our technique is compatible with additional experimental probes beyond optics and transport and opens up new avenues for future experiments in vdW heterostructure devices.

\section{Experimental methods}

Our experimental design combines the well-established piezoelectric-based strain technique for bulk samples\cite{hicks2014piezoelectric, park2020rigid, bartlett2021relationship} with the standard dry-transfer technique for fabricating vdW heterostructure devices.\cite{kim2016van} We first discuss the challenges inherent to integrating these two techniques, and overview our solution of creating a bowtie-shaped silicon substrate.

\subsection{Current challenges for straining 2D vdW devices}
vdW heterostructure devices are typically fabricated on 500~$\mu$m thick Si/SiO$_2$ wafer substrates.\cite{kim2016van,cao2018correlated,Wang20142Dfab} A simple way to induce strain to the vdW heterostructure is to directly strain the substrate wafer. However, single crystalline silicon is a rigid material, with Young's modulus of $\approx130-180$~GPa.\cite{hopcroft2010young} Hence, the force required to strain a millimeter size wafer to $\approx1$\% is approximately 750 N, far exceeds what can be achieved in existing piezo-based strain devices. Therefore, it remains an open challenge to identify a suitable substrate that is flexible enough to achieve high levels of the strain, yet rigid enough to be compatible with the fabrication of state-of-the-art vdW heterostructure devices. One approach is to replace the silicon wafer with a flexible substrate, such as a thin metal or a polymer. \cite{wang2019situ,mohiuddin2009uniaxial,wu2014piezoelectricity,zhang2017strain,ma2020piezoelectricity} For example, large tensile and compressive strains were previously induced in a vdW material by using a two- or three-point bending geometry with flexible substrates such as polyimide-coated phosphor bronze or soft polyethylene-based thin films. \cite{wang2019situ,zhang2017strain,ma2020piezoelectricity} However, bending the substrate can additionally induce an unintentional vertical strain gradient in multilayer vdW heterostructures. 

Developments of piezoelectric strain cell technology over the past decade have significantly advanced experimental capabilities for uniaxial strain tuning of bulk crystals. \cite{hicks2014piezoelectric} The key innovation was the arrangement of three parallel piezo stacks, which compensates for the large (and inverse) thermal contraction of the piezo stacks. This geometry further enables the application of much larger strain values than previously possible (up to $\approx\pm$1\%) owing by the counter motion of the central and outer piezo stacks. This strain cell is now commercially available, and it has been successfully integrated with different experimental probes including nuclear magnetic resonance,\cite{pustogow2019constraints} x-ray scattering,\cite{malinowski2020suppression,sanchez2021transport} AC specific heat,\cite{li2020heat} and the elastocaloric effect.\cite{ikeda2019ac} This development naturally leads to the possibility of using the piezoelectric strain cell to apply uniaxial strain to a silicon substrate.

However, the very high Young's modulus of silicon still presents a significant challenge. Without any modification, inducing a one percent strain in a regular silicon wafer requires a force that far exceeds the blocking force of the typical piezo stacks used in commercial strain cells. Several approaches have been developed so far in an attempt to circumvent this issue. One approach is to suspend a sample across a micrometer size gap ($\sim$3-5 $\mu$m), which is created by carefully cleaving the silicon wafer.\cite{cenker2022reversible} Although this approach can result in the application of strain exceeding 1\% to a thin vdW flake, it is extraordinarily difficult to fabricate complex vdW device architectures that are suspended across the gap. 

Another approach is to transfer the vdW onto a thin silicon wafer strip, ideally only a few hundred micrometers wide. The small cross section of the thin strip greatly reduces the amount of force required to strain the substrate. \cite{cenker2022reversible, hwangbo2023strain} However, the extremely small size of thin strip makes the fabrication of conventional vdW heterostructure devices very difficult. To address this issue, we developed a new approach in which we first shape the silicon wafer into a bowtie shape. This geometry is inspired by a similar titanium platform structure used in Refs.~\onlinecite{bartlett2021relationship} and \onlinecite{park2020rigid} to strain bulk crystals. The key advantage of this design is the narrow bridge of the bowtie greatly reduces the amount of force required to induce strain in that region, while the large area at two ends of the wafer simultaneously enables the fabrication of arbitrarily complex vdW heterostructure devices with top and bottom gates and many electrical contacts.

\begin{figure}[b]
    \centering
    \includegraphics[width = 3.5 in]{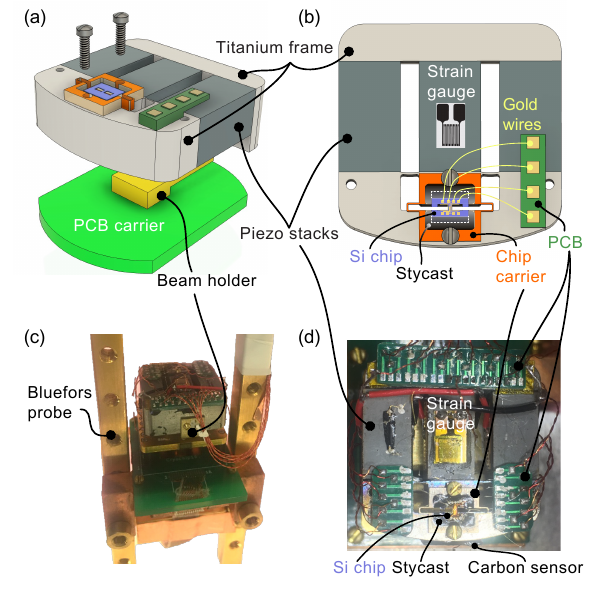}
    \caption{(a) Schematic of the strain apparatus assembly. The light gray frame is constructed of titanium. The piezo stacks (dark gray) have dimensions of $5\times5\times9$ mm.\cite{piezo} A PCB is affixed onto the side of the strain cell on the titanium frame and hovers above the nearby piezo stack. A second PCB is affixed onto the other side of the strain cell, but is not shown here for clarity. The PCB is used to connect electrical wires to both the sample and the wiring of the cryostat insert. The strain cell is affixed to a PCB carrier by a beam holder, and the entire assembly can be mounted onto various cryostat inserts.
    (b) Top view of the apparatus. The chip carrier (orange) is tightly affixed onto the titanium frame by two screws. The bowtie-shaped 50 $\mu$m thick silicon chip (purple) is epoxied onto the chip carrier with Stycast 2850FT epoxy (black). The top surface is partially enveloped by the epoxy, denoted by the white dashed lines. The 25 $\mu$m gold wires (yellow) are bonded to the PCB pads (green) using silver paste. A commercial strain gauge is glued on the central piezo stack.
    (c) Side-view photograph of the strain cell attached to a Bluefors dilution refrigerator top-loading insert.
    (d) Top-view photograph of the strain cell. A temperature sensor made of carbon composition resistor \cite{samkharadze2010new} was adhered to the sidewall of the titanium block with GE vanish to monitor the local temperature of the strain cell in the cryostat.}
    \label{fig:1}
\end{figure}

\subsection{The design of strain cell for vdW 2D device}
\begin{figure*}
    \centering
    \includegraphics[width = 7 in]{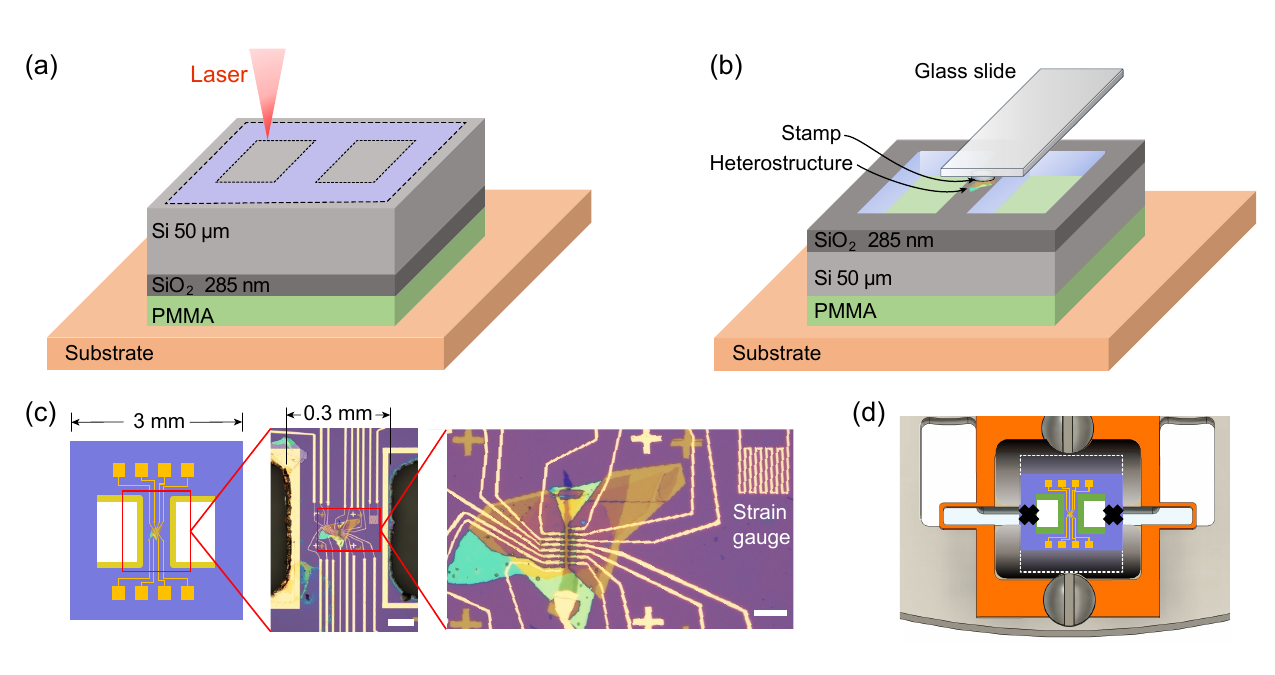}
    \caption{(a) Cartoon schematic illustrating the laser cutting process used to create a 50~$\mu$m thick bowtie-shaped silicon chip. The thin silicon wafer (gray) is first glued onto a thicker silicon substrate wafer with PMMA. The 285 nm thick SiO$_2$ capping layer faces downward to avoid overheating during laser cutting. By carefully adjusting the power, the laser can cut throughout the Si/SiO$_2$ layers without degrading the SiO$_2$ surface. The dashed lines indicate the laser cutting path. The light blue indicates that regions of the silicon wafer are removed after cutting. The outer square frame is used to isolate many bowtie-shaped chips from a single large-area wafer. 
    (b) Schematic of a standard dry transfer process of a vdW heterostructure onto the clean SiO$_2$ layer, located at the center bridge of the pre-cut silicon chip, as shown in (a). 
    (c) Layout of the final silicon chip after electron beam lithography, plasma etching, and metal deposition (left). Optical image of the central bridge region (middle), and the vdW heterostructure device with a meandering gold pattern strain gauge (right). The central bridge width is about 0.3 mm. The white scale bars represent 100 $\mu$m (middle) and 25 $\mu$m (right), respectively.
    (d) Diagram of the silicon chip glued on top of the chip carrier, which is pre-fixed by two screws on the strain cell. The epoxied area regions are outlined by the white dashed lines. The two outside arms (marked by the black crosses) are cut after the epoxy has cured.
}
    \label{fig:fab}
\end{figure*}

We use a home-built piezoelectric strain cell with a design similar to the one introduced by Hicks et al.\cite{hicks2014piezoelectric} As depicted in Fig.~\ref{fig:1}, the device comprises three piezo stacks ($5\times5\times9$ mm, with high stiffness constant and blocking force\cite{piezo}) and custom-machined titanium frames. We further include a beam holder that affixes the strain cell to a printed circuit board (PCB) sample mount, which allows us to easily wire up the strain cell for use in different cryostats (we have so far used a PPMS from Quantum Design, Inc., and a top-loading dilution refrigerator from Bluefors\cite{thermallink}). Viewed from above the strain cell, the bowtie-shaped silicon chip (colored purple) is glued with Stycast epoxy\cite{stycast} onto a titanium chip carrier (colored orange), which is prefixed onto the titanium sample mount by screws on each end. The glue covers roughly half of the top of the tab area of the bowtie to ensure robust mechanical anchoring, while leaving enough space for around eight electrical contacts on each side. The gold electrical pads are wired to a custom PCB that is mounted on the side of the strain cell. A commercial strain gauge\cite{straingauge} is glued onto one of the piezo stacks and is used to estimate the displacement of the titanium chip carrier by the piezo stacks. A small carbon resistor is adhered to the titanium frame of the strain cell in order to monitor the local temperature\cite{samkharadze2010new}.

We fabricate our vdW heterostructure devices atop a 50~$\mu$m thick silicon wafer with a 285~nm SiO$_2$ capping layer.\cite{IWS} This wafer is ten times thinner than those conventionally used for vdW heterostructure devices. The reduced thickness is important for lowering the spring constant and enabling the application of larger strain.\cite{cenker2022reversible} We first laser-cut two rectangular holes out of the wafer in order to create the bowtie shape. In contrast to previous methods used for laser cutting titanium and quartz plates\cite{park2020rigid, bartlett2021relationship}, additional steps are needed to protect the surface of SiO$_2$ layer when laser cutting the silicon wafer since the vdW heterostructure is extremely sensitive to the surface cleanliness. Therefore, prior to the cutting, we glue the thin wafer atop a thicker silicon substrate using polymethyl methacrylate (PMMA) as the bonding adhesive. The SiO$_2$ surface of the thin wafer is arranged to face downward, in contact with the PMMA adhesive. There are two benefits of doing this: first, to provide mechanical stability for the fragile thin wafer, and second, to protect the SiO$_2$ layer from silicon dust generated during the laser cutting process. The substrate wafer is affixed onto an aluminum plate with double-sided tape, and the plate is mounted on the laser cutting board. We employed a laser cutting system (LPKF ProtoLaser) to cut the silicon wafer. We note that the laser power is critical since over cutting can also damage the SiO$_2$ layer or result in poor detachment from the PMMA. Here, we use a laser power of 1.2 watts with a beam diameter of 10 $\mu$m and 1000 times cut repetition for each pattern. 
The square silicon wafer is cut into the designated pattern outlined by the dashed lines shown in Fig.~\ref{fig:fab}(a). Note that this pattern includes two outer bridges in addition to the central bridge of the bowtie, as shown in Fig.~\ref{fig:fab}(b). These outer bridges are temporary, but they are essential for providing mechanical stability during the transfer and fabrication of vdW heterostructure device. These two outer bridges are cut by hand after mounting the wafer onto the piezoelectric strain cell. The bridges of the bowtie are aligned along <110> directions of a (100) wafer. 

After laser cutting, the small three-bridge-patterned thin silicon chips are detached from the Si substrate wafer by sonicating in acetone. The three-bridge-patterned chip is then glued to another rigid silicon substrate with PMMA, this time with its SiO$_2$ layer facing upwards [Fig. \ref{fig:fab}(b)]. A vdW heterostructure is assembled layer-by-layer using the standard dry-transfer technique with a poly-carbonate (PC) stamp. After the stack is assembled, the PC stamp is melted onto the central bridge of the thin silicon wafer and soaked in chloroform to dissolve the PC, leaving behind only the vdW heterostructure. Since the patterned wafer detaches from the thicker Si/SiO$_2$ substrate wafer in chloroform, it must be re-affixed to a new rigid substrate for further processing. The vdW heterostructure is then processed into a Hall bar geometry and contacted electrically following a standard series of steps involving electron beam lithography, plasma etching, and metal deposition. The bowtie pattern created in the silicon wafer does not impede PMMA spin coating and electron-beam lithography; thus, these procedures are essentially the same as in the standard fabrication process for vdW heterostructures. We arrange the Hall bar such that it is aligned along the direction of uniaxial strain. Figure \ref{fig:fab}(c) shows a typical three-bridge chip with a tBLG device at its center. In addition to the gold electrodes connected to the device, we also evaporate gold along the laser-cut edges of the chip. These gold strips are essential; otherwise, the rough wafer edges resulting from laser cutting can prevent a smooth metal liftoff procedure. Additionally, the roughness can make the wafer more prone to breaking under tensile strain, but these can be smoothed and potentially strengthened through chemical etching. 
We also evaporate a meander-shaped gold wire near the device, which serves as a custom local strain gauge. The three-bridge silicon chip is then epoxied to the titanium chip carrier pre-mounted on the strain cell, as described above. The final step is to cut the outer two bridges [at positions indicated by the black crosses in Fig.~\ref{fig:fab}(d)] using a scalpel blade after the epoxy has cured.  

\section{Strain calibration}
\begin{figure}
    \centering
    \includegraphics[width = 3.3 in]{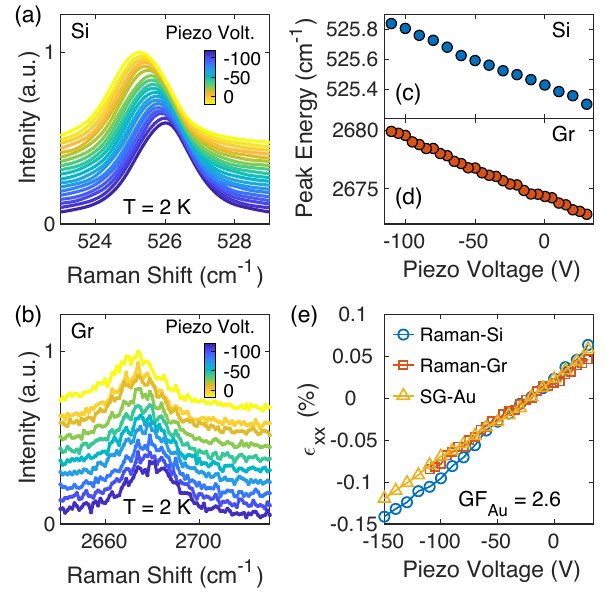}
    \caption{(a) and (b) Raman spectra of the bowtie-shaped silicon and graphene at different piezo voltages, acquired at a temperature of 2 K.  
    (c) and (d) The energy of the Raman peak of silicon and graphene as a function of piezo voltage, indicating a linear relationship.
    (e) Measured strain as a function of piezo voltage. The negative piezo voltage corresponds to the direction of compressive strain. The strain values were estimated from the position of the Raman peaks in (c) and (d) or determined by the gold SG deposited on silicon and the commercial SG glued on the piezo stack. The zero-strain point was calibrated by measuring the Raman spectrum at a strain-free spot, located far from the central bridge of the Si wafer. The Raman peak is at 525.5 cm$^{-1}$, corresponding to a piezo voltage of nearly -20 volts. The gauge factor of the gold SG is obtained from the quadratic fitting in Fig.~\ref{fig:gold}(b). The strain values calculated by the SG match well with those measured by Raman spectroscopy, with all exhibiting a linear dependence on the piezo voltage. 
    }
    \label{fig:Raman}
\end{figure}

\begin{figure*}
    \centering
    \includegraphics[width = 6.5 in]{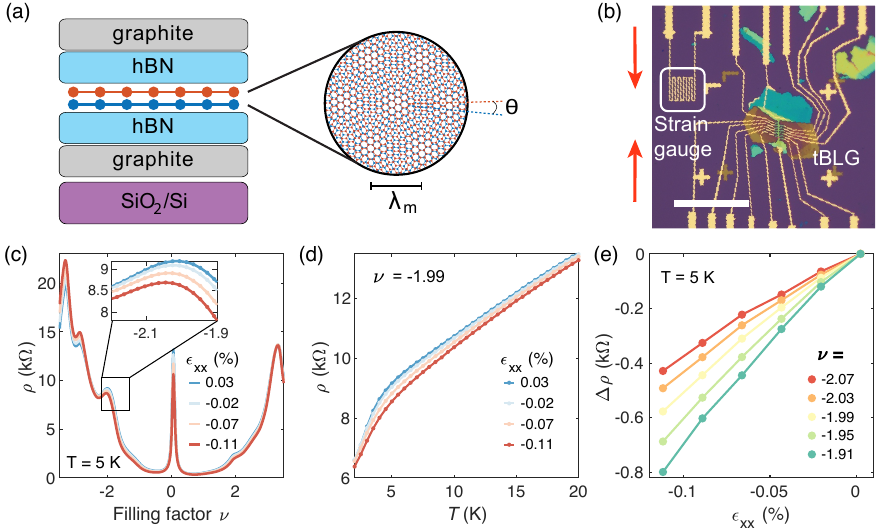}
    \caption{(a) Schematic of a twisted bilayer graphene sample, encapsulated in hBN with graphite gates top and bottom gates. The right panel displays a cartoon illustration of the moiré pattern of tBLG. The moiré wavelength is $\lambda_{\rm m} =a/[2\sin(\theta/2)]$, where $a=0.246$ nm is the lattice constant of graphene and $\theta$ represents the twist angle.
    (b) Optical image of a tBLG device shaped into a Hall bar geometry resting on a bowtie-shaped silicon chip. The long axis of the Hall bar is aligned with the uniaxial strain direction, indicated by the red arrows. A nearby gold strain gauge, with a line width of 2 $\mu$m, is highlighted in the white box. The white scale bar is 100 $\mu$m. 
    (c) Resistivity vs band filling factor at different strains for a tBLG device.\cite{tseng2022anomalous} Inset: Zoom-in view of the resistivity nearby $\nu = -2$. 
    (d) Temperature dependence of resistivity at different strains. 
    (e) The relative resistivity changes as a function of strain at several filling factors, showing linearity with strain.
}
    \label{fig:tBLG}
\end{figure*}

In most piezoelectric strain cell experiments, the strain level of the sample was estimated by measuring the displacement of piezo stacks using a capacitance strain gauge.\cite{park2020rigid,bartlett2021relationship} The measured displacement was then converted to the strain of the sample by using a strain transmission factor or an effective length determined by the finite element analysis. We performed similar measurements and analysis and present the results in Appendix \ref{app_FEA}. We have also developed two methods to directly measure the amount of strain induced in the vdW heterostructure as we bias the piezo stacks. We discovered that these two methods provide more accurate and reliable measurement of the strain, which we discuss below.

 In the first method, we perform Raman spectroscopy on both the silicon wafer and on a monolayer graphene encapsulated in hBN. We use a laser power of 300 $\mu$W, with integration times of 160 seconds for graphene and 10 seconds for silicon. These Raman spectra allow us to precisely determine the magnitude of uniaxial strain independently of graphene and silicon using the relationships previously established in Refs.~\onlinecite{wang2019situ} and \onlinecite{urena2013raman}. Figures~\ref{fig:Raman}(c) and \ref{fig:Raman}(d) show the Raman spectra of silicon substrate and graphene taken at different piezo voltages at a temperature of 2 K. Both peaks systematically blueshift due to the effect of compressive uniaxial strain, as shown in Figs.~\ref{fig:Raman}(e) and \ref{fig:Raman}(f). We note that the strain cell applies uniaxial stress, which induces both symmetric ($A_{1g}$ irrep) and antisymmetric ($E_g$ irrep, in the case of graphene) components of strain. The linear shift of Raman peak energy is due to the symmetric component of strain. The in-plane Poisson’s ratio along <110> directions of a (100) silicon is only 0.064, indicating a nearly uniaxial strain along the stress direction. This uniaxial strain condition is the same as the Raman measurement performed in Ref. \onlinecite{urena2013raman} (Si) and \onlinecite{wang2019situ} (graphene), hence we can reliably use the data in previous work to extract the strain value. 
 We convert the wavenumber of the Raman peak to uniaxial strain following the previously established relationships and plot the measured strain value as a function of piezo voltage in Fig. \ref{fig:Raman}(e). We note that a higher bias up to $300$~V, if allowed by the instrumental setup, can be safely applied to the piezo stack at cryogenic temperatures and would likely induce a larger strain. \cite{hicks2014piezoelectric,park2020rigid} The strain induced in the graphene and silicon both vary linearly as a function of piezo voltage, and overlap almost perfectly. The slightly reduced strain in graphene compared with the silicon indicates that $\approx 80$\% of the strain induced in the silicon wafer is transferred to graphene. 

Although Raman spectroscopy provides the most direct means of calibrating the strain in the vdW heterostructure, it is often not possible to perform \textit{in situ} optical measurements. For example, transport measurements performed in a dilution refrigerator are generally incompatible with Raman characterization. Therefore, we have also explored electrical means of calibrating the applied strain. The most common technique is to measure the resistance of a long metal wire as the piezo bias is tuned. We have developed custom strain gauges that we fabricate by evaporating a gold meander nearby the vdW heterostructure, as shown in Fig.~\ref{fig:fab}(c). The measured elastoresistance can be converted to strain by 
\begin{equation}
    \epsilon = \frac{\Delta R/R}{GF},
    \label{eq1}
\end{equation}
where $GF$ is the gauge factor of the strain gauge. There are, however, two sources of uncertainty inherent to this technique. First, the strain is measured in silicon and not directly in the vdW material of interest. Second, the gauge factor of the gold strain gauge can vary between $\approx2.1$ and $3.3$ in our experiments, based on the precise details of the device fabrication (more details can be found in Appendix \ref{app_gold}). Nevertheless, we observed a systematic variation of the gauge factor as a function of the gauge resistivity. By fitting the gauge factor vs resistivity with a quadratic function, we estimated the gauge factor of the gold strain gauge evaporated near the same device measured by the Rama spectroscopy.  Figure \ref{fig:Raman}(e) shows the measured strain from the gold strain gauge compared with the strain measured by the Raman spectroscopy. We see that both methods are in good agreement. Therefore, the evaporated gold strain gauges provide a reliable rough estimate of the induced strain in the vdW heterostructure and can be used to confirm that the induced strain in the wafer is linear in the applied piezo voltage.

\section{The effect of strain on transport in a twisted bilayer graphene device}

In order to demonstrate the power of our strain technique, we measure the strain-dependent transport of a vdW heterostructure device consisting of a twisted bilayer graphene device. The twist angle of the device is $\theta=1.2^\circ$, very near the magic angle at which superconductivity and other correlated states are prominent. The rich correlated and topological physics of magic-angle tBLG has been explored in great detail elsewhere;\cite{balents2020superconductivity,andrei2020graphene} here, we use it as a model system to test our technique, given that the band structure is expected to be highly sensitive to strain. The device consists of tBLG encapsulated by flakes of hBN, all resting on a flake of graphite (further details on this device can be found in Ref.~\onlinecite{tseng2022anomalous}). The graphite acts as a back gate and can change the charge carrier density in the tBLG when a voltage is applied between the graphite and tBLG. Figure~\ref{fig:tBLG}(c) shows a measurement of the longitudinal resistivity, $\rho$ of the device as a function of the band filling factor, $\nu$ ($\nu=4$ corresponds to fully filled moir\'e bands, where the factor of four reflects the spin and valley degeneracy of graphene). In our measurements performed at a temperature of 5 K, we see a resistive state at $\nu=0$ indicative of the charge neutrality point, as well as resistive states at other selected integer values of $\nu$ indicative of incipient correlated insulating states. As we apply uniaxial strain by changing the bias on the piezo stacks, we see that the resistivity of the device can change by hundreds of ohms, depending on the precise value of $\nu$.

The effect of strain is especially prominent around $\nu=-2$, which corresponds to a developing correlated insulating state at half filling of the moir\'e valence band in the tBLG. Figure~\ref{fig:tBLG}(d) shows how $\rho$ evolves both as a function of temperature and strain at a fixed band filling factor of $\nu=-1.99$. Overall, the temperature dependence is characteristic of tBLG devices previously reported.\cite{cao2020strange,polshyn2019large,jaoui2022quantum} By comparing $\rho(T)$ at different values of applied strain, we see that there is a monotonic decrease of the resistivity at all temperatures up to 20 K as compressive strain is applied. Figure~\ref{fig:tBLG}(e) shows the relative resistivity, $\Delta \rho$, as a function of compressive strain at values of $\nu\approx-2$. In all cases, we see a nearly linear change in resistivity as strain is applied. Although future work is necessary to unravel the physics underlying the strain-tuning of these states, the large linear elastoresistance we see clearly illustrates the ability of strain to manipulate the electronic properties of tBLG.

\section{Conclusion}

In summary, we have reported the development of a new technique for applying continuously tunable strain to vdW heterostructure devices of ultrahigh quality and arbitrary complexity. The strain device is based on a commercially available three piezo stack design, with custom PCBs added in order to electrically connect to vdW heterostructure devices. We achieved a large strain in excess of $-0.15$\% by appropriately modifying the silicon substrate. We are able to demonstrate efficient strain transfer from the piezo stacks to the vdW heterostructure. As shown in Fig. \ref{fig:Raman}(e), the measured strain is linear in the applied piezo voltage over the entire range we studied. This linearity implies that the total strain we can achieve can very likely be increased by further improvements in epoxying both the top and bottom surfaces of the silicon wafer to the sample mount, and by applying larger voltages to the piezo stacks up to $300$~V. Our technique paves the way for future characterization and control of the correlated and topological physics of a variety of vdW heterostructures. The technique is readily compatible with a range of other characterization tools, including scanning probe microscopes.

\section*{acknowledgments}
This work was mainly supported by the NSF MRSEC at UW under Grant Number DMR-2308979, the Gordon and Betty Moore Foundation’s EPiQS Initiative, grant GBMF6759 to J.-H.C. and the U.S. Department of Energy, Office of Science, Office of Basic Energy Sciences under Award Number DE-SC0023062. The vdW heterostructure device fabrication was partially supported by the Army Research Office under Grant Number W911NF-20-1-0211. J.-H.C. acknowledges support from the David and Lucile Packard Foundation. M.Y. and J.-H.C. also acknowledge support from the State of Washington-funded Clean Energy Institute. This work made use of shared fabrication facilities provided by NSF MRSEC 1719797. 

\section*{AUTHOR DECLARATIONS}

\subsection*{Conflict of Interest}
The authors have no conflicts to disclose.

\subsection*{Author contributions}
Zhaoyu Liu and Xuetao Ma contributed equally to this paper.\\


\appendix

\section{Strain estimation via finite element analysis}\label{app_FEA}
\begin{figure}[t]
    \centering
    \includegraphics[width = 3.5 in]{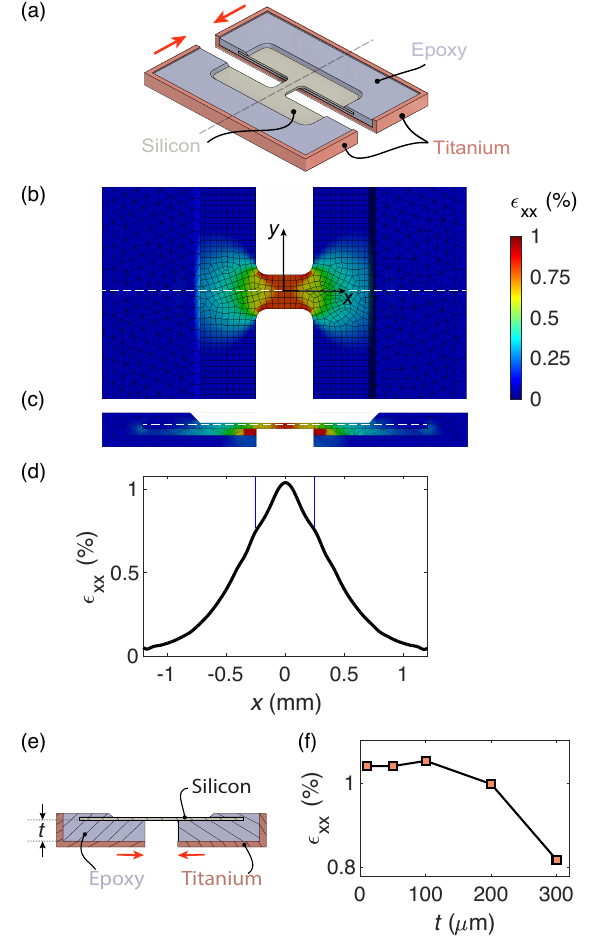}
    \caption{(a) Illustration of a bowtie-shaped silicon chip for FEA. The chip is encapsulated with an epoxy layer (50 $\mu$m thick). The red arrows indicate the moving directions of the titanium plates. The strain distribution in a top-down view (b) and a cross-sectional view along the dashed centerline (c).  
    (d) Strain variations along a designated path, marked by a dashed line in (c). Blue lines highlight the gap region.
    (e) The cross-sectional view along the dashed line in (a) exhibits the model characterized by thickness denoted as \textit{t}.
    (f) Plot strain at the central point in (d) as a function of the epoxy thickness.
    }
    \label{fig:FEA}
\end{figure}

In addition to the direct measurement of the strain in the silicon substrate and graphene as discussed in the main text, we also estimate the induced strain by measuring the displacement of the piezo stacks and perform finite element analysis. To do this, we epoxy a commercial strain gauge directly atop one of the piezo stacks [Fig.~\ref{fig:1}(b) and (d)]. Since the gauge factor of the commercial strain gauge is well calibrated, we can directly extract the strain in the piezo stacks, $\epsilon_{\rm piezo}$, and the displacement of the piezo stacks, $L_{\rm piezo}\times\epsilon_{\rm piezo}$, where $L_{\rm piezo}=9~$mm is the length of the piezo stack. The displacement allows us to estimate the strain in the silicon substrate, $\epsilon_{\rm xx}^{\rm norm}$, by the following relationship:
\begin{equation}
    \epsilon_{\rm xx}^{\rm norm}= \mu\frac{2L_{\rm piezo}}{L_{\rm gap}}\epsilon_{\rm piezo},
    \label{eq2}
\end{equation}
where $L_{\rm gap} = 0.5~$mm is the distance between the titanium sample holder and $\mu$ is a dimensionless strain transmission factor that takes into account the strain relaxation due to the epoxy and within the silicon substrate. The factor of 2 takes into account the displacement of both the center piezo stacks and the outer two piezo stacks. 

To estimate the strain transmission factor $\mu$, we performed FEA using ANSYS Academic Research Mechanical software to investigate the strain distribution in the bowtie-shaped silicon chip model, as shown in Fig.~\ref{fig:FEA}(a). Young's modulus and Poisson's ratio of the silicon along the <110> directions of a (100) wafer in the simulation are 169 GPa and 0.064, respectively. \cite{hopcroft2010young} Young’s modulus of the SiO$_2$ layer is 70 GPa. The narrow bridge of the silicon chip is 0.3 mm in width and 50 $\mu$m in thickness. Using $k = Y\frac{A}{L}$, we estimated the effective spring constants of Si and SiO$_2$ layers on a strip substrate matching the dimensions of the central bridge of the bowtie-shaped chip. The effective spring constant of the silicon substrate is $k = 5.1~N/\mu m$, which is significantly larger that of the SiO$_2$ layer, $0.012~N/\mu m$. Hence, the SiO$_2$ layer can be safely ignored in our FEA simulation.

In this simulation, a 5 $\mu$m displacement was applied to both sides of the titanium plate, which results in a total 2\% nominal uniaxial strain. We assume the thickness of the epoxy layer to be 50 $\mu$m. In Fig. \ref{fig:FEA}(b, c), we present both the top and cross-sectional views of the strain distribution. Notably, the strain primarily aligns along the white dashed center-line of the silicon wafer. The highest strain region (colored in red) is located at the center, where the vdW heterostructure was fabricated. 
As shown in Fig. \ref{fig:FEA}(d), in the simulation the strain at the center of the silicon bridge is 1.04\%, resulting in a strain transmission factor $\mu_{\rm FEA}= 52\%$. Nevertheless, this simulation does not take into account that the titanium strain cell itself may also deform. Using the effective spring constant of $12~N/\mu m$ for the strain cell\cite{park2020rigid} and $5.1~N/\mu m$ for the substrate, we estimate that only 70\% of the displacement induced by the piezo actuator is transferred to the silicon substrate. Taking this factor into account, the strain estimated by this method is within a factor of two of the strain measured by the Raman spectroscopy [Fig.~\ref{fig:Raman}(e)].  

Additionally, we simulate the strain at the center of the bridge as a function of the epoxy thickness, $t$. The results, presented in Fig.~\ref{fig:FEA}(f), demonstrate that the applied strain gradually decreases as the epoxy thickness exceeds 100~$\mu$m. However, in practice, achieving a precise thickness of epoxy presents a challenge. We attempt to minimize the thickness by using as little epoxy as possible and by carefully pressing the wafer down in order to extrude excess epoxy out from underneath.

\section{Gold strain gauge calibration}
\label{app_gold}
\begin{figure}[tb]
    \centering
    \includegraphics[width = 3.4 in]{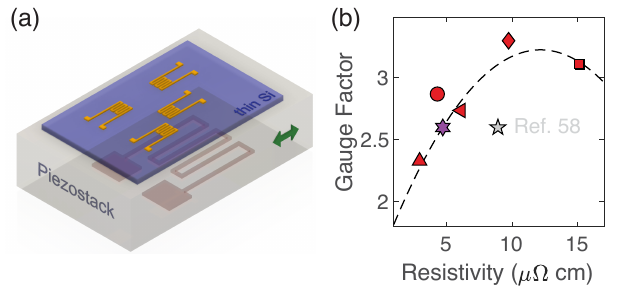}
    \caption{(a) Configuration for measuring the gold strain gauges with a single piezo stack technique. The Si/SiO$_2$ substrate and commercial SG were glued to top and bottom of the piezo stack with epoxy, respectively. The gold SGs were directly deposited on SiO$_2$ surface. The green arrow presents the poling direction of the piezo stack.
    (b) Gauge factor of the gold SG as a function of the resistivity at 2 K. The different red symbols refer to different measurements. The purple hexagram indicates the gold SG in Fig.~\ref{fig:Raman}(e). 
    The data can be fitted by a quadratic curve (dashed line): $\rm GF=-0.0113\rho^2+0.2753\rho+1.553$}. 
    \label{fig:gold}
\end{figure}

To calibrate the gold SGs, we have conducted measurements on multiple gold SGs using a single piezo stack technique. \cite{shayegan2003low} The gold SGs were fabricated either directly on top of a Si/SiO$_2$ substrate or above several single-hBN flakes, as shown in Fig. \ref{fig:gold}(a). The silicon wafer is glued to one side of the piezo stack, while a commercial SG is affixed on the other side. At low temperature, the strain can be almost fully transferred to the silicon wafer which can be precisely measured by the commercial SG.\cite{chu2012divergent} After determining the strain of the silicon wafer, we calibrate the gold SGs by measuring the change of resistance of the gold SGs fabricated on silicon wafer and obtain the gauge factor using Eq.~(\ref{eq1}). In Fig.~\ref{fig:gold}(b), we plot the obtained gauge factor as a function of the resistivity of the gold SGs. The gauge factor ranges from 2.1 to 3.3, in agreement with the previous report,\cite{li1994thin} and follows a quadratic behavior as a function of resistivity. The fitted quadratic function allows us to determine the gauge factor of gold SGs in other experiments. We have also estimated the strain transmission through the vdW heterostructures by taking the ratio between the strain measured from gold SGs on hBN flakes and those on the silicon wafer. The ratio is typically above 90\% and remains nearly independent of temperature and thickness below 20 K. While the variation of the gauge factor between different gold SG does not allow us to directly interpret this ratio as the actual strain transmission factor, this is in broad agreement with the results of Raman measurement presented in Fig. \ref{fig:Raman} in the main text.

\bibliography{strain2D}

\begin{thebibliography}{58}
\bibitem [{the()}]{thermallink}%
  \BibitemOpen
  \href@noop {} {}\bibinfo {note} {An additional thermal link, made of copper tape, will be used to connect the titanium block to the probe in the dilution refrigerator}\BibitemShut {NoStop}%
\end{thebibliography}

\end{document}